\documentclass{article}
\usepackage{arxiv}
\usepackage{amssymb,amsmath,stmaryrd}
\usepackage{amsthm}
\usepackage{algorithm}
\usepackage[noend]{algpseudocode}
\usepackage{float}
\usepackage{graphicx}
\usepackage{xcolor}
\usepackage{latexsym}
\usepackage{hyperref}
\usepackage{subcaption}

\usepackage{todonotes}

\usepackage{complexity}

\graphicspath{{figures/}{charts/}}

\newcommand{\etal}{\textit{et al}.}

\newtheorem{lemma}{Lemma}
\algnewcommand\algorithmicforeach{\textbf{for each}}
\algdef{S}[FOR]{ForEach}[1]{\algorithmicforeach\ #1\ \algorithmicdo}
\renewcommand{\G}{\mathcal{G}}
\renewcommand{\H}{\mathcal{H}}
\DeclareMathOperator{\argmax}{argmax}

\newcommand{\overbar}[1]{\mkern 3mu\overline{\mkern-3mu#1\mkern-3mu}\mkern 3mu}

\def\probat(#1,#2,#3){\mathcal{P}[#1, #2; #3]}
\def\probtill(#1,#2,#3){\mathcal{P}[#1, #2; \preccurlyeq #3]}
\def\edgeprob(#1,#2){p_{#1 \shortrightarrow #2}}
\def\paths(#1,#2){\mathcal{P}_{#1 \leadsto #2}}
\def\qaths(#1,#2){\mathcal{Q}_{#1 \leadsto #2}}
\def\prob[#1]{\mathrm{Prob}[#1]}
\def\influencedby(#1,#2){X_{#1 \triangleright #2}}
\def\qinfluencedby(#1,#2){Y_{#1 \triangleright #2}}
\def\qwinfluencedby(#1,#2,#3){Y_{#1 \triangleright #2 \shortrightarrow #3}}
\def\sinfluencedby(#1,#2){Z_{#1 \triangleright #2}}
\def\ninfluencedby(#1,#2,#3){X_{\overline{#1}, #2 \triangleright #3}}
\def\qninfluencedby(#1,#2,#3){Y_{\overline{#1}, #2 \triangleright #3}}
\newcommand{\maxinfl}{\mathit{MaxInfluence}}
\newcommand{\infl}{\mathit{Influence}}

\newcommand{\state}[1]{#1.\mathit{state}}
\newcommand{\level}[1]{#1.\mathit{level}}
 
\DeclareMathOperator{\last}{last} 
 
\renewcommand{\O}[1]{\mathcal{O}(#1)}

\makeatletter

\newsavebox\myboxA

\newsavebox\myboxB

\newlength\mylenA
\newcommand*\xoverline[2][0.75]{%

	\sbox{\myboxA}{$\m@th#2$}%

	\setbox\myboxB\null

	\ht\myboxB=\ht\myboxA%

	\dp\myboxB=\dp\myboxA%

	\wd\myboxB=#1\wd\myboxA

	\sbox\myboxB{$\m@th\overline{\copy\myboxB}$}

	\setlength\mylenA{\the\wd\myboxA}

	\addtolength\mylenA{-\the\wd\myboxB}%

	\ifdim\wd\myboxB<\wd\myboxA%

	\rlap{\hskip 0.5\mylenA\usebox\myboxB}{\usebox\myboxA}%

	\else

	\hskip -0.5\mylenA\rlap{\usebox\myboxA}{\hskip 0.5\mylenA\usebox\myboxB}%

	\fi}

\makeatother
\title{Maximizing the Influence: Analytic Activation Probability Computation Approach}

\author{
	Maryam Adineh \\
	Department of Computer Engineering\\ Ferdowsi University of Mashhad\\ Mashhad, Iran \\
	\texttt{maryam.adineh@mail.um.ac.ir} \\
	\And
	Mostafa Nouri-Baygi\thanks{Corresponding author} \\
	Department of Computer Engineering\\ Ferdowsi University of Mashhad\\ Mashhad, Iran \\
	\texttt{nouribaygi@um.ac.ir} \\
}

\begin{document}
\maketitle

\section*{Abstract}
Influence maximization is the problem of finding a subset of the most influential individuals in a network. The impact of social networks on the dissemination of information and the development of viral marketing has made this problem as the subject of many studies. Influence maximization is $\NP$-hard and many greedy algorithms have been proposed to solve the problem.

In this paper we propose a greedy algorithm that approximates the influence by using a novel analytic activation probability computation method. We propose a nonlinear system of equations to compute the activation probabilities which is more accurate than state-of-the-arts.
Moreover, we propose a method to compute the activation probabilities efficiently, in order to reduce the running time of the algorithm. 

We examine our proposed methods on some real-world data sets.
Our experiments demonstrate that the proposed algorithms outperform the previous works by computing more accurate activation probabilities. In addition, Our efficient method has much improvements on the running-time, so it can be easily scaled on large networks.

\textbf{Keywords} Activation probability computation, Heuristic algorithms, Independent cascade model, Influence maximization, Influence propagation
\section{Introduction}
Nowadays, social networks have provided a good platform for disseminating information, ideas and innovations. Dissemination of information on social networks is done by word-of-mouth effect which plays an important role in introducing and promoting products. One popular type of social advertising namely viral marketing takes advantage of this effect.

Consider a newly established company that develops an application and wants to promote it to all individuals in a social network. One way is to advertise the application directly which will be so costly. In addition, users would not trust them easily. Another way is to select a small subset of the most influential individuals who are interested in the application and giving it to them by some gifts. Consequently, these individuals would encourage their friends to buy it and the friends would encourage their friends' friends and so on. Finally the product will be adopted by a large population of the network. 

The problem of finding the most influential individuals  which is known as the \emph{Influence Maximization} problem was first introduced by Domingos and Richardson~\cite{domingos2001mining,richardson2002mining}. Kempe \etal~\cite{kempe2003maximizing} proved that the influence maximization problem is $\NP$-hard. Also they proposed a hill climbing greedy algorithm for the problem and proved that this greedy algorithm approximates the solution within $63\%$ of the optimal solution for independent cascade model and linear threshold model. The hill climbing algorithm uses Monte~Carlo simulation to estimate the influence spread of each vertex. Since the simulation is too time-consuming, many studies have been carried out in an attempt to develop new algorithms with less running time.

\subsection{Our contributions}
In this paper, we propose two greedy heuristics to solve the problem. We approximate the influence spread by using the activation probability computation method. 
We propose a nonlinear system of equations to compute the activation probabilities which is more accurate than state-of-the-arts. Moreover, we propose a method to compute the efficient activation probabilities in order to reduce the running time. 

The main idea of our methods is filling the gap between approximated and exact value of activation probability computation.
Our experiments show that the proposed algorithms outperform the previous methods by propagating more influence spread. In addition, Our efficient method has much improvements on the running-time so it can be easily scaled on large networks.

The rest of the paper is organized as follows. first we discuss the related work in Section~\ref{sec-related}. In Section~\ref{sec-problem} we describe the problem definition. We propose our methods in Section~\ref{sec-proposed}. Experimental results are evaluated in Section~\ref{sec-experiments} and we conclude the paper in Section~\ref{sec-conclusion}.
\section{Related Work}\label{sec-related}

Influence maximization problem has been extensively studied and many heuristics are proposed to solve this problem. For the first time, Kempe \etal~\cite{kempe2003maximizing} formulated the problem and proved that it is $\NP$-hard. They proposed a greedy hill climbing algorithm which guarantees the solution with $1-1/e-\epsilon$ factor of the optimal solution on the independent cascade and linear threshold models. Since Monte-Carlo simulations are used to estimate the influence of vertices, their algorithm is very time-consuming. As a result, it is not efficient on large-scale social networks.

In order to improve the efficiency of the hill climbing algorithm,  Leskovec \etal~\cite{leskovec2007cost} proposed Cost-Effective Lazy Forward (CELF) optimization which reduces the number of influence spread estimations and according to their experimental results it is 700 times faster than greedy hill climbing algorithm.

Chen \etal~\cite{chen2009efficient} proposed a new greedy algorithm to improve the running-time of previous greedy algorithms. Also they proposed a degree centrality heuristic named \emph{degree discount} which is very fast and efficient even on large-scale social networks but does not guarantee the accuracy of the solution.

Kimura \etal~\cite{kimura2006tractable} proposed a new Shortest Path Model (SPM) considering just the most efficient information spread in the network. In the SPM model, each vertex just has a chance to be activated from the seed set through the shortest path between them.

Chen \etal~\cite{chen2010scalableMIA} with this idea that the most influence of each vertex is obtained locally, proposed PMIA algorithm with the concept of maximum influence path. PMIA computes the influence of each vertex in its local region, so it is less time consuming but PMIA doesn't consider influence spread of many paths which affect its accuracy. Moreover since PMIA stores the local region of each vertex, its memory usage is too high.

Jung \etal~\cite{jung2012irie} proposed a ranking influence method namely IRIE that reduces the running-time of heavy influence estimations and consumes less memory than PMIA. 

Cheng \etal~\cite{cheng2014imrank} proposed IMRank which finds a self-consistent ranking to estimate the marginal influence of each vertex from an initial ranking.

Tang \etal~\cite{tang2018efficient} illustrated the relationship of influence spread and number of hops between the vertices.
They showed that when the number of hops between the vertices increases, the amount of influence propagates between them decreases. Therefore, most of influence propagation happens in the first hops. Consequently, they introduced OneHop and TwoHops algorithms which calculate the influence of each vertex after one and two hops, respectively. Although these algorithms will be more efficient in terms of running-time, since influence spread of many paths is ignored, the accuracy of their algorithms is low. Furthermore, for longer influence propagation paths, new formula must be developed which makes the methods less flexible.

Aggarwal \etal~\cite{aggarwal2011flow} developed \emph{steady~state} algorithm which computes the activation probaility of the vertices using a system of nonlinear equations. Then the influence spread of each vertex is estimated to select a set of most influential vertices. 

Yang \etal~\cite{yang2012approximation} proved the convergence of steady~state method and investigated its structural defect problem which causes inaccuracy in computed activation probabilities. To solve this problem and with the idea of PMIA they proposed a shortest path based algorithm which computes the activation probabilities just through the shortest paths between the vertices. Since they don't consider the activation probabilities of different paths created by cycles, this algorithm is not very effective in approximating the influence.

To reduce the structural defect problem of the steady~state method, Yang \etal~\cite{yang2019influence} proposed SSS-Noself algorithm. Although this algorithm reduces the amount of computation error, it is very time-consuming and is not efficient even on small social networks.

To improve the running time of greedy hill climbing, Borgs \etal~\cite{borgs2014maximizing} proposed a reverse sampling method which generates reverse reachable sets from randomly selected nodes. Then the seed set is selected greedily from them. Since their algorithm has some limitations in generating samples, Tang \etal~\cite{tang2014influence,tang2015influence} proposed \emph{TIM} and \emph{IMM} to overcome its drawbacks and improve the running time.

Hung \etal~\cite{nguyen2016stop} proposed \emph{SSA} and \emph{DSSA} that reduce the running time of IMM by reducing the number of randomly generated sets.
\section{Problem Definition}\label{sec-problem}

We consider a social network as a directed graph $\G(V,E)$ that $|V|=n$ and $|E|=m$. Each directed edge $(u,v)$ has a propagation probability $\edgeprob(u,v) \in [0,1]$ which is the probability that $v$ will be activated by $u$. In order to calculate the amount of influence that propagates through the graph starting from a subset of nodes, we use the \emph{independent cascade model (ICM)}, one of the basic diffusion models. A node is called \emph{active} if it has been influenced.

In the IC model, when a node $u$ becomes active at step $t$, it tries to activate its inactive neighbor $v$ with probability $\edgeprob(u,v)$ at step $t+1$. Each activated vertex just has one chance to activate each of its neighbors. The diffusion process terminates when no more activation is possible.
 
Given a graph $\G$, constant $k$ and the diffusion model $M$, the influence maximization problem is to find a subset of vertices $S \in V$ such that their influence spread, denoted by $\sigma(S)$, maximizes.
\section{Proposed Methods}\label{sec-proposed}

Consider a non-negative, monotone and submodular function $f$ that $f(\emptyset) = 0$.  Nemhauser \etal~\cite{nemhauser1978analysis} showed that if we select a set $S$ of size $k$ that maximizes $f(S)$ according to Algorithm~\ref{greedy}, the algorithm will guarantee $f(S) > (1 - 1 / e) f(S^*)$ that $S^*$ is the optimal solution.
A function $f$ is called \emph{submodular}, if for sets $S\subseteq T \subseteq V$ and any $v\in V\backslash T$, the marginal gain of adding $v$ to $S$ is not smaller than the marginal gain of adding it to $T$, that is, $f(S\cup \{v\})-f(S)\geq f(T\cup \{v\})-f(T)$. A function $f$ is called to be \emph{monotone} if $f(S)\leq f(T)$.

\begin{algorithm}
	\caption{Greedy hill climbing}\label{greedy}
	\begin{algorithmic}[1]
		\Require $\G(V,E)$: social network graph; $k$: size of the result set; $f$: monotone and submodular set function.
		\Ensure $S$: seed set.
		
		\State $S\gets \emptyset$
		\ForAll {$i=1$ to $k$}
		\State $u$=$\argmax_{w\in V\backslash S}(f(S\cup\{w\})-f(S))$
		\State $S\leftarrow S\cup\{u\}$
		\EndFor
		\State \Return{ $S$}
	\end{algorithmic}
\end{algorithm}

Kempe \etal~\cite{kempe2003maximizing} expressed that influence spread function $\sigma(\cdot)$ is monotone and submodular and used the greedy algorithm which selects $k$ seeds with the most influence spread, iteratively. They estimated the influence of each vertex by simulating the influence spread for several thousand times which is very time-consuming. As explained in Section~\ref{sec-related}, later papers proposed other approaches based on the activation probability computation to calculate the influence.

The main idea of our proposed algorithms is to calculate the influence of each vertex by computing the activation probability of the vertices. Furthermore, we propose an approach to compute the activation probabilities efficiently in order to reduce the running-time. 

The difference betweeb our method and other methods based on activation probability computation is that, the old methods determine the activation probabilities for the long-run steady state. This way of calculation overestimates the activation probability of vertices by propagating the influence of a vertex to itself through cycles, which we call the \emph{resonance effect}. The resonance effect is much significant specially when the graph is highly connected.

In AACPA and EAAPC, to compute the activation probability of a vertex, we consider the probability of not being activated before. Furthermore, we compute the activation probabilities until an appropriate step $T$, so the computation error produced by the resonance effect is reduced and probabilities are computed more accurately. As will be shown by the results of Section~\ref{sec-experiments}, the activation probability which is computed in AACPA and EAACPA is more accurate than previous methods.

\subsection{Influence Computation}\label{sec_approx_influence}
Let $S$ be a set of seeds and ICM be the influence diffusion model.
We describe the method on directed graphs but it can be expanded to undirected graphs simply by replacing each undirected edge with two directed edges.
For a vertex $v$, $\probat(S,v,t)$ represents the \emph{activation probability} of $v$ at step $t$ given initial seed set $S$, that is, the probability of $v$ to become activated by $S$ at step $t$. Similarly, $\probtill(S,v,t)$ represents the activation probability of $v$ \emph{until} step $t$ that is the probability of $v$ to be activated at step $t$ given seed set $S$. Lemma~\ref{lemma_recurrence_relation} explains how to calculate these two probabilities using recurrence relations for each vertex.  

\begin{lemma}\label{lemma_recurrence_relation}
Let $S$ be an initial seed set. For each vertex $v$, the activation probability at step $t$ and the activation probability until step $t$ can be calculated according to the following recurrence relations.

\begin{eqnarray}
\probat(S, v, t) & = & (1 - \probtill(S, v, t-1)) \cdot \nonumber\\[1ex]
&   & \qquad \big(1 - \prod_{u: (u,v) \in E} (1 - \edgeprob(u, v) \cdot \probat(S, u, t-1))\big),\label{eq_prob_at} \\[2ex]
\probtill(S, v, t) & = & 1 - \prod_{\tau=0}^{t} (1-\probat(S, v, \tau)).\label{eq_prob_till}
\end{eqnarray}
And for initial step $t=0$ we have
\begin{equation}
\probat(S, v, 0) = \probtill(S, v, 0) = \left\{
\begin{array}{rl}
1 & \text{if $v \in S$} \\
0 & \text{otherwise.}
\end{array}\label{eq_prob_base}
\right.
\end{equation}

\end{lemma}

\begin{proof}
A vertex $v$ is active at step $t$, if it has become activated in any step $\tau=0, \cdots, t$. The probability of this event can be calculated by its complementary event that $v$ is not active at $t$, which means it has not been activated in any steps $\tau=0, \cdots, t$. Therefore,
$$1 - \probtill(S, v, t) = \prod_{\tau=0}^{t} (1-\probat(S, v, \tau)),$$
which proves Equation~\ref{eq_prob_till}.

In this equation, the events of not being activated at each steps are assumed independent. Hence, the probability of not being activated at any of these times is equal to the product of them.

It is clear that, $v$ becomes active at step $t$, if it has not been activated until $t-1$ and at least, one of its in-neighbors which became active at step $t-1$ succeeds to activate $v$. This probability is exactly equal to Equation~\ref{eq_prob_at}.

At step $t=0$, since the vertices have not affected each other yet, the activation probability is $1$ for seeds and $0$ for the other vertices. So the Equation~\ref{eq_prob_base} is proved.
\end{proof}

The recurrence relations of Lemma~\ref{lemma_recurrence_relation} are obtained assuming that the events are independent. 
As will be explained in the next subsection, this assumption is not precisely true, but in graphs with the characteristics of social networks the difference with real probabilities will be very small.  

With these recurrence relations, we can simply calculate the activation probability of each vertex at step $t$ and the activation probability until step $t$. The activation probability of vertex $v$ is $\probtill(S, v, \infty)$, but, as the value of $t$ increases, the changes to $\probtill(S, v, t)$ will become negligible, until the probability values reach the steady state. The maximum appropriate value for $t$ depends on the features of the graph, but as we will show in section~\ref{sec-exper-accuray}, in most applicable graphs it is enough to calculate the probabilities until $T=6$.

After formulating the activation probability of each vertex, we show how to find the influence spread of a seed set $S$. Lemma~\ref{lemma_influence} describes how to calculate the influence spread of a seed set according to the activation probability of the vertices. 

\begin{lemma}\label{lemma_influence}
Let $S$ be an initial seed set. The influence spread of $S$ is the following
$$\sigma(S) = \sum_{v \in V} \probtill(S, v, \infty).$$
\end{lemma}
\begin{proof}
	Let $X_v$ be a random variable whose value is equal to $1$, if $v$ is activated at the end of the influence propagation process, and $0$ otherwise. We have
	
	\begin{eqnarray}
	\sigma(S) &=& E\big[|\{v | v \text{ is active} \}|\big] \nonumber\\[1ex]
	&=& E\big[\sum_{v \in V} X_v\big] \nonumber\\[1ex]
	&=& \sum_{v \in V} E[X_v] \label{eqn_sum_of_expected}\\[1ex]
	&=& \sum_{v \in V} \probtill(S, v, \infty).\nonumber
	\end{eqnarray}
	
	Equation~\ref{eqn_sum_of_expected} follows from the linearity of expectation.
\end{proof}

According to Lemma~\ref{lemma_influence}, we can estimate the influence spread of the seed set $S$ by summing over the activation probabilities of the vertices at a specified time which is big enough. In subsequent experiments, it will be shown that the appropriate value is $T=6$.

With this equations we now propose our greedy algorithm called AAPC.

\subsection{Analytic Activation Probability Computation Algorithm}\label{sec_AAPC}
As stated before, Kempe \etal's~\cite{kempe2003maximizing} greedy hill climbing estimates the influence by running simulations several thousand times. This causes the algorithm to be very time consuming. In addition, its running-time is dependent on the value of propagation probabilities between the vertices, that is, the higher the propagation probabilities, the more running time.

As explained in Section~\ref{sec_approx_influence}, we can approximate the influence of each seed set by summing over the activation probabilities of vertices of the graph. We here describe a greedy algorithm, Analytic Activation Probability Computation or AAPC for short, using this technique. Our proposed algorithm starts from an initial empty seed set. At every step, we calculate the marginal influence gained by adding each vertex to the current seed set and add the one with the maximum marginal influence to the seed set. Algorithm~\ref{alg_approx} gives the details of AAPC.

\begin{algorithm}
	\caption{AAPC}\label{alg_approx}
	\begin{algorithmic}[1]
		\Require $\G(V,E)$: social network graph; $k$: size of the result set; $p$: propagation probability of edges in the ICM.
		\Ensure $S$: seed set.
		
		\State $S\gets \emptyset$
		\For {$\mathit{iter} \gets 1 \text{ to } k$}
		\State $\maxinfl \gets 0$
		\ForEach {$w \in V\backslash S$}
		\State $S \gets S \cup \{w\}$ \Comment{Temporarily add $w$ to the seed set}
		\ForEach {$v \in V$}
		\If {$v \in S$}
		\State{$\probat(S, v, 0) \gets \probtill(S, v, 0) \gets 1$}
		\Else
		\State{$\probat(S, v, 0) \gets \probtill(S, v, 0) \gets 0$}
		\EndIf
		\EndFor
		\For {$t \gets 1$ to $T$} \Comment{Find the probability of activation for each $t$}
		\ForEach {$v \in V$}
		\State $\probat(S, v, t) = (1 - \probtill(S, v, t-1)) \cdot$
		\Statex $\qquad\qquad\qquad\qquad\qquad\qquad \big(1 - \prod_{u: (u,v) \in E} (1 - \edgeprob(u, v) \cdot \probat(S, u, t-1))\big)$
		\State $\probtill(S, v, t) = 1 - \prod_{\tau=0}^{t} (1-\probat(S, v, \tau))$
		\EndFor
		\EndFor
		\State $\infl \gets \sum_{v \in V} \probtill(S, v, T)$ \Comment{Sum over all probabilities}
		\If {$\infl > \maxinfl$}
		\State $\maxinfl \gets \infl$
		\State $\mathit{seed}_{\mathit{iter}} \gets w$ 
		\EndIf
		\State $S \gets S\backslash \{w\}$
		\EndFor
		\State $S \gets S \cup \{\mathit{seed}_{\mathit{iter}}\}$
		\EndFor
		\State \Return {$S$}
	\end{algorithmic}
\end{algorithm}

Initially, the propagation probability of all the vertices are set to 0. Then, we select a vertex $w$ as the only seed and calculate the activation probability until $T$ for any vertex $v$, that is $\probtill(\{w\}, v, T)$. According to Lemma~\ref{lemma_influence}, summation of these activation probabilities gives the influence spread of $w$. So the vertex with the highest influence spread is added to the seed set. In the next steps, the marginal influence of adding each vertex to the seed set will be calculated in a similar way and the vertex with the maximum marginal influence will be selected as the next seed. This process continue until $k$ vertices are selected. 

\subsubsection{Analysis of AAPC}
AAPC has a considerable improvement against greedy algorithm of Kempe \etal~\cite{kempe2003maximizing} in terms of the running time. The time complexity of AAPC is $\O{kTnm}$, and with the assumption that $T<10$ is a small constant, the time complexity can be written as $\O{knm}$. However the time complexity of Kempe \etal's~\cite{kempe2003maximizing} is $\O{kRnm}$, while $R$ is a big constants, such as $10000$. 

Despite this improvement, however, AAPC is not still applicable for very large-scale social networks with millions or billions of vertices. Applying the idea of CELF~\cite{leskovec2007cost} on AAPC may reduce the running time in practice.

Another issue in AAPC is that we assume the activation probabilities are independent, while it is not precisely true. This causes the influence spread calculated by Equation~\ref{eq_prob_at} and \ref{eq_prob_till} to be more than the exact values. Indeed, when there are too many cycles in the graph, the influence of each vertex returns to itself, and it makes the calculated influence to be greater than the exact value. 

\begin{figure}
	\begin{center}
\begingroup%
  \makeatletter%
  \providecommand\color[2][]{%
    \errmessage{(Inkscape) Color is used for the text in Inkscape, but the package 'color.sty' is not loaded}%
    \renewcommand\color[2][]{}%
  }%
  \providecommand\transparent[1]{%
    \errmessage{(Inkscape) Transparency is used (non-zero) for the text in Inkscape, but the package 'transparent.sty' is not loaded}%
    \renewcommand\transparent[1]{}%
  }%
  \providecommand\rotatebox[2]{#2}%
  \newcommand*\fsize{\dimexpr\f@size pt\relax}%
  \newcommand*\lineheight[1]{\fontsize{\fsize}{#1\fsize}\selectfont}%
  \ifx\svgwidth\undefined%
    \setlength{\unitlength}{150.4414937bp}%
    \ifx\svgscale\undefined%
      \relax%
    \else%
      \setlength{\unitlength}{\unitlength * \real{\svgscale}}%
    \fi%
  \else%
    \setlength{\unitlength}{\svgwidth}%
  \fi%
  \global\let\svgwidth\undefined%
  \global\let\svgscale\undefined%
  \makeatother%
  \begin{picture}(1,0.47776571)%
    \lineheight{1}%
    \setlength\tabcolsep{0pt}%
    \put(0,0){\includegraphics[width=\unitlength,page=1]{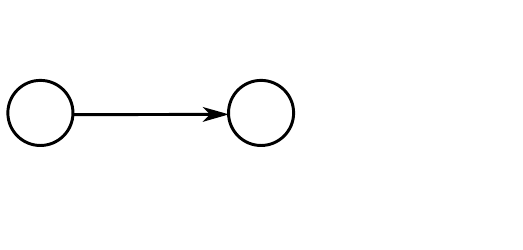}}%
    \put(0.05790249,0.2381239){\color[rgb]{0,0,0}\makebox(0,0)[lt]{\lineheight{1.25}\smash{\begin{tabular}[t]{l}$u$\end{tabular}}}}%
    \put(0,0){\includegraphics[width=\unitlength,page=2]{cycle.pdf}}%
    \put(0.89789653,0.0455696){\color[rgb]{0,0,0}\makebox(0,0)[lt]{\lineheight{1.25}\smash{\begin{tabular}[t]{l}$x$\end{tabular}}}}%
    \put(0,0){\includegraphics[width=\unitlength,page=3]{cycle.pdf}}%
    \put(0.90813415,0.39414992){\color[rgb]{0,0,0}\makebox(0,0)[lt]{\lineheight{1.25}\smash{\begin{tabular}[t]{l}$w$\end{tabular}}}}%
    \put(0.48042334,0.24809457){\color[rgb]{0,0,0}\makebox(0,0)[lt]{\lineheight{1.25}\smash{\begin{tabular}[t]{l}$v$\end{tabular}}}}%
    \put(0,0){\includegraphics[width=\unitlength,page=4]{cycle.pdf}}%
  \end{picture}%
\endgroup%

	\end{center}
	\caption{In this graph, there is a cycle: $v \to w \to x \to v$.}\label{pic_cycle}
\end{figure}

For example, in the graph of Figure~\ref{pic_cycle}, assume that the propagation probability for all edges is a constant value $p$ and vertex $u$ is selected as the only seed. The exact value for the activation probability of $v$, $w$ and $x$ are $p$, $p^2$ and $p^3$, respectively.

We now compute the activation probability of the vertices of the graph according to Lemma~\ref{lemma_recurrence_relation}, as the following. In the first step, $t=0$, $u$ is only activated and its activation probability is equal to $1$, i.e., $\probat(\{u\},u,0)=1$, and the activation probability of the other vertices is zero. In step $t=1$, the activation probability of all vertices are calculated. Only, $v$ will have non-zero probability and will become active by $u$ with probability $p$. Since $v$ is activated in step $t=1$, it can activate $w$ in step $t=2$ and we will have $\probat(\{u\},w,2)=p^2$. 

In the same way, $x$ will be activated by $w$ in $t=3$ and its activation probability will be equal to $p^3$. Since a neighbor of $v$ has non-zero activation probability in step $t=3$, in step $t=4$, once more $v$'s neighbor has non-zero activation probability, So according to Equation~\ref{eq_prob_at}, $\probat(u,v,4)=(1-p)p^4$. This happens while $x$ is activated by $w$ and $w$ is activated by $v$. Consequently, the activation probability of $w$ at $t=2$ and the activation probability of $x$ at $t=3$ are not independent of the activation probability of $v$ at $t=1$.

Since the activation probabilities are computed in several steps, further computations in such a cyclic graph, will cause the activation probability of a vertex to return back to itself. Therefore, the computed activation probabilities will sometimes be more than the exact values.

In this example, it is interesting to explain the difference between our method and previous probability-based methods. As mentioned before, in the prior methods, the activation probability is calculated in the steady state, which means the activation probability of $v$ in the previous steps is ignored. So the activation probability of $v$ after four steps is $p + p^4(1-p)$ while in our method this probability is equal to 
$$1 - (1-p)[1-(1-p)p^4] = p + p^4(1-p)^2.$$

It is clear that, the exact value of the activation probability for $v$ is $p$ and any additional value is computational error of the method. In this step of the algorithm, the previous methods have $p^4(1-p)$ extra value while our method adds less value which is $p^4 (1-p)^2$. In this step, the additional value calculated for $v$ is less than which is calculated in the previous methods and as the number of steps increases, it will still be smaller. In addition, by increasing the value of $p$ to get close to $1$ the difference between the computational error in our method and the previous methods will grow.

Using this example, we conclude that in probability-based methods if the number of steps in which the probability is computed is bigger (that is, larger $T$), the number of traversed cycles will be more and it will increase the amount of influence that returns to each vertex. But, if the number of steps is small, the returned influence through many cycles is negligible.

Furthermore, if the number of cycles in a graph is small or cycles are very long, the amount of returned influence to each vertex will be negligible in this algorithm. On the other hand, if the propagation probability on an edge is high, the returned influence to a vertex will increase. However, as explained in the example, the latter is more significant in previous methods and in ours, the negative effect of increasing the probability is smaller or it may even have desirable effect.

Even though the running-time of Algorithm~\ref{alg_approx} has been considerably improved compared to the greedy algorithm of Kempe \etal~\cite{kempe2003maximizing}, it is still time consuming on large-scale graphs. In the next section we propose \emph{Efficient Analytic Activation Probability Computation} algorithm which is very faster by considering just the most efficient activation probabilities. We refer to this algorithm as EAAPC.

\subsection{EAAPC Algorithm}\label{sec_EAAPC}

Some probability-based algorithms, including \cite{chen2010scalableMIA,yang2012approximation}, in order to reduce the running-time, calculate the activation probability just through a shortest path. In spite of their small running time, since they avoid to consider the activation probabilities that propagate through most paths, they are not accurate enough. Moreover, the algorithm by Chen \etal~\cite{chen2010scalableMIA} saves local regions for all vertices and thus has high memory overhead.

In EAAPC, similar to AAPC, we compute the influence of all vertices according to their activation probability propagation, and then select the vertex with the most influence as a seed. The main difference in this algorithm is how to calculate the activation probabilities. 

To reduce the running time and the effect of the returned influence, the main idea of EAAPC is to compute the probabilities for effective paths, that is, the probabilities propagate through all shortest paths and ignore insignificant probabilities. 

We here, first, describe how to select the suitable paths and calculate the probabilities through these paths when there is only one seed, and then, we generalize it for the case that there are more than one seed.

In the IC model, each activated vertex tries to activate its neighbors through the outgoing edges. As Kempe \etal~\cite{kempe2003maximizing} expressed with the \emph{Triggering Model}, it is equivalent to assume that each vertex selects a subset of its incoming edges namely the \emph{Triggering Set} randomly and only is influenced through the selected edges.

By this equivalence, we can generate a random graph $\H$ from $\G$ such that their set of vertices are the same and each edge $(u, v) \in E$ can be in $\H$ with probability $\edgeprob(u,v)$. Given $\H$ and seed set $S$, the influence of $S$, $\sigma(S)$, is the expected number of reachable vertices from $S$ and the activation probability of vertex $v$ is the probability of being reachable from at least one of the vertices in $S$.

Assume $\influencedby(u, v)$ is the event of $v$ becoming activated by $u$. Let $\paths(u,v)$ denote the set of all paths from $u$ to $v$ in $\G$. We say that, path $\pi$ is in $\H$, if all edges of $\pi$ exist in $\H$. If some path of $\paths(u,v)$ is in $\H$, $v$ is influenced by $u$. Thus the probability of $v$ becoming activated by $u$ is equal to
$$\prob[\influencedby(u, v)] = \prob[\exists \pi \in \paths(u, v) \colon \pi \in \H].$$

The probability of a path to be in $\H$ can be calculated simply by $$\prob[\pi \in \H] = \prod_{(u, v) \in \pi}\edgeprob(u, v).$$ But it is not easy to calculate $\prob[\exists \pi \in \paths(u, v): \pi \in \H]$ due to the large number of paths and (possible) non-empty intersections between each pair of paths.
 
To solve this problem, we sort paths from $u$ to $v$ in an arbitrary order. Among all paths from $u$ to $v$ in graph $\H$, the shortest path propagates the influence to $v$. Thus we sort all the paths in non-descending order of their length.
Assume the ordering is $\paths(u,v) = \langle \pi_1, \pi_2, \dots, \pi_l \rangle$, where $l$ is the number of paths in $\paths(u,v)$. The activation probability of $v$, which is propagated by $u$ is calculated using this ordering, according to Lemma~\ref{lemma_bfs_u_influence}.

\begin{lemma}\label{lemma_bfs_u_influence}
Let $\paths(u, v) = \langle \pi_1, \pi_2, \dots, \pi_l \rangle$ be a list of all paths from $u$ to $v$ in $\G$, sorted in non-decreasing order of length. The activation probability of $v$ from $u$ is
$$\prob[\influencedby(u, v)] = \sum_{1 \le i \le l} \prob[\pi_i \in \H \land \forall j < i, \pi_j \not\in \H]$$
\end{lemma}

\begin{proof}
The activation probability of $v$ from $u$ is equal to the probability of existing a path from $u$ to $v$ in $\H$. Let $A_\pi$ be the event that $\pi$ is in $\H$, then
$$\prob[\influencedby(u,v)] = \prob[\cup_{\pi \in \paths(u,v)} A_\pi].$$

Based on the given ordering, let $B_\pi$ be the event that $\pi$ is in $\H$ but none of its predecessors are. It is clear that 
$$B_\pi = A_\pi \cap \left (\cap_{\pi' \in \paths(u,v): \pi' \prec \pi} \overline{A}_{\pi'} \right ).$$
In this equation, $\pi' \prec \pi$ means that in the given ordering, $\pi'$ is before $\pi$ and $\overline{A}_{\pi'}$ is the event that $\pi'$ is not in $\H$. Event $B_\pi$ discards parts of $A_\pi$ which are covered by $A_{\pi'}$. Therefore, for each two paths $\pi$ and $\pi'$, $B_\pi$ and $B_{\pi'}$ are disjoint, and
$$\cup_{\pi \in \paths(u,v)} A_\pi = \cup_{\pi \in \paths(u,v)} B_\pi.$$

To calculate the probability of occurrence of at least one of the events $A_\pi$ we can write
\begin{eqnarray}
\prob[\cup_{\pi \in \paths(u,v)} A_\pi] & = & \prob[\cup_{\pi \in \paths(u,v)} B_\pi] \label{eqn_prob_union_bs} \\
& = & \sum_{\pi \in \paths(u,v)} \prob[B_\pi] \label{eqn_sum_prob_bs} \\
& = & \sum_{\pi \in \paths(u,v)} \prob[(A_\pi \cap (\cap_{\pi' \in \paths(u,v)}\overline{A}_{\pi'}))]\nonumber \\
& = & \sum_{\pi_i \in \paths(u,v)} \prob[\pi_i \in \H \land (\forall j < i, \pi_j \not\in \H)].\nonumber
\end{eqnarray}

Equation~\ref{eqn_sum_prob_bs} follows from Equation~\ref{eqn_prob_union_bs} since the events are disjoint.
\end{proof}

Lemma~\ref{lemma_bfs_u_influence} suggests a method to calculate the activation probability of the vertices given only one vertex as a seed. However, there is a difficulty here. There are too many paths, which may be exponential in terms of the number of vertices, and make the probability computation impossible in a polynomial time. In the following we describe a method to calculate the activation probability in a polynomial time with a small error.

Select a subset of paths from $u$ to $v$ as the following set
\begin{eqnarray}
\qaths(u,v) = \{ \, \pi \in \paths(u, v) & \mid & \forall \pi' \in \paths(u,v) \colon \label{eqn_qath}\\
&&(\pi' \neq \pi \to  d_\pi(u, \last(\pi, \pi')) \le d_{\pi'}(u, \last(\pi, \pi'))) \,\}.\nonumber
\end{eqnarray}
In the above equation, $\last(\pi, \pi')$ denotes the last common vertex between two paths $\pi$ and $\pi'$ before their common end-vertex $v$, and  $d_\pi(u, w)$ denotes the number of edges from $u$ to $w$ in $\pi$. Since $u$ is the starting vertex of both $\pi$ and $\pi'$, $\last(\pi, \pi')$ is always defined.

According to Equation~\ref{eqn_qath}, $\qaths(u, v) \subseteq \paths(u, v)$ selects paths from $u$ to $v$ that are shortest among all paths from $u$ to the last common vertex before $v$. It can be assured that a significant portion of the influential paths in activating $v$ are included in $\qaths(u, v)$ because a path $\pi'$ is not added to $\qaths(u,v)$ if there is a similar path $\pi$ from $u$ to $v$ that has less traversed edges from $u$ to the last common vertex before $v$. Since the number of traversed edges in $\pi$ from $u$ to $\last(\pi, \pi')$ is less than the number of traversed edges of $\pi'$, the probability of $\last(\pi, \pi')$ becoming activated by $u$ via $\pi$ is highly greater, and so removing $\pi'$ does not significantly effect the activation probability.

Using this result, the idea is to compute the probability of activation of $v$ by $u$ only through paths of $\qaths(u,v)$. We define $\qinfluencedby(u,v)$ as the event that one of the paths from $\qaths(u,v)$ is in $\H$. We have
$$\prob[\qinfluencedby(u,v)] = \prob[\exists \pi \in \qaths(u,v): \pi \in \H].$$
According to the above justification, probability $\prob[\qinfluencedby(u,v)]$ is very close to $\prob[\influencedby(u,v)]$.

We here compare the definition of $\qaths(u,v)$ with the Maximum Influence Path defined by Wang \etal~\cite{wang2012scalable}, in which \emph{only one} path is considered as the maximum influential path.
Many paths with equal probabilities may be the most influential path but in MIP only one of them is considered. There may be many paths that have slightly smaller or even greater probabilities, but they are not considered in MIP.

When we choose $\qaths(u,v)$ as the established paths, we are sure that all paths to $v$ from a given neighbor $w$ have equal number of edges from $u$ to $w$, and similarly from $u$ to $v$.

Let $\qwinfluencedby(u,w,v)$ be the event of $v$ becoming activated by $u$ via $w$ through one of the existent paths in $\qaths(u,v)$ and $\sinfluencedby(u,w)$ be the event of $w$ becoming activated by $u$ via one of the shortest paths between $u$ and $w$. 
$\qwinfluencedby(u,w,v)$ occurs if at least one of the paths from $\qaths(u,v)$ that pass through $w$ be in $\H$. Thus
$$\prob[\qwinfluencedby(u, w, v)] = \prob[\exists \pi \in \qaths(u, v) \colon w \in \pi \land \pi \in \H].$$

In a similar way, $w$ can be activated through one of the shortest paths from $u$ to $w$ that is in $\H$. Thus

\begin{multline}
\prob[\sinfluencedby(u, w)] = \prob[\exists \pi \in \paths(u, w) \colon \pi \in \H \land \\ \forall \pi' \in \paths(u, w),\ d_\pi(u, w) \le d_{\pi'}(u, w)]. \qquad\quad\nonumber
\end{multline}

Since all paths in $\qaths(u,v)$ which cause the event $\qwinfluencedby(u,w,v)$ to be occurred are shortest paths from $u$ to $w$, we can write
$$\prob[\qwinfluencedby(u,w,v)] = \prob[\sinfluencedby(u,w)] \cdot \edgeprob(w,v).$$

Vertex $v$ can not be activated if none of its neighbors are activated, so by using complementary probability we can write

\begin{eqnarray}
\prob[\qinfluencedby(u,v)] & = & 1 - \prod_{w: (w, v) \in E} \left(1 - \prob[\qwinfluencedby(u,w,v)]\right) \nonumber \\
& = & 1 - \prod_{w: (w, v) \in E} \left(1 - \prob[\sinfluencedby(u,w)] \cdot \edgeprob(w,v)\right). \label{eq_qinf}
\end{eqnarray}

If we know the value of $\prob[\sinfluencedby(u,w)]$ for each vertex $w$, we can calculate $\prob[\qinfluencedby(u,v)]$ using the above equations. The value of $\prob[\sinfluencedby(u,w)]$ is equal to the probability that at least one of the shortest paths from $u$ to $w$ is in $\H$. It is calculated simply by the equation $$\prob[\sinfluencedby(u,w)] = \probat(\{u\},w,t),$$
where $t$ is the length of the shortest path from $u$ to $w$. Using Equation~\ref{eq_prob_at} in Lemma~\ref{lemma_recurrence_relation}, with the assumption that $t$ is the length of the shortest path from $u$ to $w$ and $\probtill(\{u\},w,t-1) = 0$, it is easy to find the value of $\probat(\{u\},w,t)$. Thus, by performing a breadth-first search (BFS) starting at $u$, we can compute $\prob[\sinfluencedby(u,w)]$ for each vertex $w$.

In addition, we are able to compute $\prob[\qinfluencedby(u,v)]$ by Equation~\ref{eq_qinf} while advancing the BFS. Algorithm~\ref{alg_y_probs} represents the pseudocode of computing $\prob[\qinfluencedby(u,v)]$ for each vertex $v$ considering $u$ as the only seed. In this algorithm a threshold level MAX\_LEVEL is selected and execution of BFS continuous until this level. It is clear that, when the distance between vertices and seed grows, the amount of changes in activation probabilities reduces and continuing execution of BFS does not change this probability significantly. 

Let $p_{\textit{avg}}$ be the expected influence propagation probability via the edges in the graph and $\epsilon$ be the tolerable error of probability computation, then MAX\_LEVEL can be obtained through the equation
$$\text{MAX\_LEVEL} = \lceil \log_{p_{\textit{avg}}} \epsilon \rceil.$$
In other word, BFS process continues until the expected influence that propagates from $u$ to a vertex is less then $p_{\textit{avg}}^\text{MAX\_LEVEL}$ which is at most $\epsilon$.

As explained before, $\prob[\qinfluencedby(u,v)]$ estimates $\prob[\influencedby(u,v)]$ to a good extent. Thus, in order to compute the influence of a vertex $u$, it is enough to run Algorithm~\ref{alg_y_probs} for $u$. Eventually, the influence spread of $u$ is equal to the total value of activation probability of all vertices according to Lemma~\ref{lemma_influence}. The vertex with the maximum influence can be selected as the initial seed.

\subsection{Generalizing to Multi-Seed State}

We explained how to compute the activation probabilities when only there is one seed. Now we expand it to the case that we have more than one seed.

For a seed set $S$, $\ninfluencedby(S,u,v)$ denotes to the event of $v$ becoming activated by $u$ and not becoming activated by any of seeds in $S$. We also generalize definition of the event $\influencedby(u,v)$ to the case that there is a set of seeds, i.e., $\influencedby(S,v)$ is the event of $v$ becoming activate by $S$.

Let $S = \langle u_1, u_2, \dots, u_k \rangle$ be a sequence of seed vertices and $S_i = \langle u_1, u_2, \dots, u_i \rangle$ be a sequence of first $i$ seeds. Following lemma provides a recurrence relation to compute  $\prob[\influencedby(S_i,v)]$ according to event $\influencedby(u_i,v)$.

\begin{lemma}\label{lemma_bfs_S_influence}
Let $S_i = \langle u_1, u_2, \dots, u_i \rangle$ be a sequence of $i$ selected seeds, then the probability of $v$ to becoming activated by this seed set is equal to
$$\prob[\influencedby(S_i,v)] = \prob[\influencedby(S_{i-1},v)] + (1 - \prob[\influencedby(S_{i-1},v)]) \cdot \prob[\influencedby(u_i,v)]$$
\end{lemma}

\begin{proof}
The event $\influencedby(S_i,v)$ occurs if $v$ has been activated by one of the vertices from $S_i$. In other words:
$$\influencedby(S_i,v) = \influencedby(S_{i-1},v) \cup \influencedby(u_i,v).$$\\

Since $\ninfluencedby(S_{i-1},u_i,v) = \influencedby(u_i,v) \backslash \influencedby(S_{i-1},v)$, 
we can write 
$$\influencedby(S_i,v) = \influencedby(S_{i-1},v) \cup \ninfluencedby(S_{i-1},u_i,v).$$

It should be noticed that, last two considered sets in the union are independent, thus
$$\prob[\influencedby(S_i,v)] = \prob[\influencedby(S_{i-1},v)] + \prob[\ninfluencedby(S_{i-1},u_i,v)].$$

Moreover, if the events of $v$ to become activated by any of seeds be independent, then the probability of $v$ to become activated by $u_i$ and not to become activated by $u_j$ for $j < i$ can be written as the product of probability of these two events as the following
$$\prob[\ninfluencedby(S_{i-1},u_i,v)] = \prob[\overbar{\influencedby(S_{i-1},v)}] \cdot \prob[\influencedby(u_i,v)]$$

From combination of last two equations with following 
$$\prob[\overbar{\influencedby(S_{i-1},v)}] = 1 - \prob[\influencedby(S_{i-1},v)]$$

this lemma is proved.
\end{proof}

Lemma~\ref{lemma_bfs_S_influence} proposes a method to compute the activation probability of each vertex propagating through the seed set while the seed set grows. Before adding $u_i$ to the seed set, it contains vertices from sequence $S_{i-1}$ and we have computed activation probabilities according to that seed set. When we add $u_i$ to the seed set, we just need to add the probability of activating $v$ by $u_i$ and not by $u_j$ ($j < i$) to its previous activation probability. This probability is equal to product of probability of not becoming activated until now and probability of becoming activated by $u_i$.

In a similar way we generalize the definition of $\qinfluencedby(u,v)$ for a seed set $S$ and denote to it by $\qinfluencedby(S,v)$. Also we define $\qninfluencedby(S,u,v)$ as the event of $v$ becoming activated by $u$ via one of the path from $\qaths(u,v)$ and not becoming activated by any of the vertices from $S$. More precisely

$$\prob[\qninfluencedby(S,u,v)] = \prob[\exists \pi \in \qaths(u,v): \pi \in \H \land \forall u' \in S \forall \pi' \in \paths(u',v): \pi' \not \in \H ].$$

For $\prob[\qinfluencedby(S_i,v)]$ we also can obtain a similar result to lemma~\ref{lemma_bfs_S_influence}. Finally we complete the algorithm of activation probability computation. Algorithm~\ref{alg_y2_probs} represents how it works. 
It is worth noting in this algorithm that, if we meet a seed like $u_j$ while traversing a path, the activation probability of $u_j$ does not increase because its previous activation probability is $1$. So $u_j$ will not change the probability of its next vertices and the traverse will not continue from $u_j$. 

\begin{algorithm}
	\caption{\textsc{ApproximateActivationProbs}($\G$, $u$)}\label{alg_y_probs}
	\begin{algorithmic}[1]
	\Require $\G=(V,E)$: the social network graph; $u$: the current candidate seed.
	\Ensure $\prob[\qinfluencedby(u,v)]$: the probability of activation for each vertex $v \in V$
	\State Set $Q$ as an empty queue.
	\ForEach {$v \in V$}
	\State $\state{v} \gets$ \textbf{unprocessed}
	\State $\prob[\overbar{\sinfluencedby(u,v)}] \gets 1$ \Comment{Probability of the complement of $\sinfluencedby(u,v)$ is initially $1$ for all $v$}
	\EndFor
	\State Enqueue$(Q, u)$ \Comment{Perform a BFS starting from $u$.}
	\State $\state{u}$ $\gets$ \textbf{in-process}
	\State $\level{u} \gets 0$
	\State $\prob[\overbar{\sinfluencedby(u,u)}] \gets 0$ \Comment{Probability of the complement of $\sinfluencedby(u,u)$ is $0$}
	\While {$Q$ is not empty}	
	\State $v \gets$ Dequeue$(Q)$
	\State $\prob[\sinfluencedby(u,v)] \gets 1 - \prob[\overbar{\sinfluencedby(u,v)}]$ \Comment{Finalize the probability of $\sinfluencedby(u,v)$}
	\State $\prob[\overbar{\qinfluencedby(u,v)}] \gets \prob[\overbar{\sinfluencedby(u,v)}]$ \Comment{Probability of the complement of $\qinfluencedby(u,v)$}
	\If {$\level{v} <$ MAX\_LEVEL}
	\ForEach {$(v, w) \in E$}
	\If {$\state{w} \neq$ \textbf{processed}}
	\If {$\state{w} = $ \textbf{unprocessed}}
	\State Enqueue$(Q, w)$
	\State $\state{w} \gets$ \textbf{in-process}
	\State $\level{w} \gets \level{v} + 1$
	\EndIf
	\State $\prob[\overbar{\sinfluencedby(u,w)}] \gets \prob[\overbar{\sinfluencedby(u,w)}] \cdot (1-\edgeprob(v,w) \cdot \prob[\sinfluencedby(u,v)]) $
	\Else
	\State $\prob[\overbar{\qinfluencedby(u,v)}] \gets \prob[\overbar{\qinfluencedby(u,v)}] \cdot (1-\edgeprob(v,w) \cdot \prob[\sinfluencedby(u,v)])$
	\EndIf
	\EndFor
	\EndIf
	\State $\state{v} \gets$ \textbf{processed}
	\EndWhile
	\ForEach {$v \in V$}
	\State $\prob[\qinfluencedby(u,v)] \gets 1 - \prob[\overbar{\qinfluencedby(u,v)}]$ \Comment{Finalize the probability of $\qinfluencedby(u,v)$}
	\EndFor
    \State \Return $\prob[\qinfluencedby(u,v)]$ for each $v \in V$
	\end{algorithmic}
\end{algorithm}

\begin{algorithm}
    \caption{\textsc{MultiSeedApproximateActivationProbs}($\G$, $S_{i-1}$, $u$)}\label{alg_y2_probs}
    \begin{algorithmic}[1]
        \Require $\G=(V,E)$: the social network graph; $S_{i-1}$: the set of previous selected seeds; $u$: the current candidate seed.
        \Ensure $\prob[\qinfluencedby(S_{i-1}\cup \{u\},v)]$: the probability of activation by $S_{i-1}\cup u$ for each vertex $v \in V$
        \State Set $Q$ as an empty queue.
        \ForEach {$v \in V$}
        \State $\state{v} \gets$ \textbf{unprocessed}
        \State $\prob[\overbar{\sinfluencedby(u,v)}] \gets 1$ \Comment{Probability of the complement of $\sinfluencedby(u,v)$ is initially $1$ for all $v$}
        \EndFor
        \State Enqueue$(Q, u)$ \Comment{Perform a BFS starting from $u$.}
        \State $\state{u}$ $\gets$ \textbf{in-process}
        \State $\level{u} \gets 0$
        \State $\prob[\overbar{\sinfluencedby(u,u)}] \gets 0$ \Comment{Probability of the complement of $\sinfluencedby(u,u)$ is $0$}
        \While {$Q$ is not empty}	
        \State $v \gets$ Dequeue$(Q)$
        \State $\prob[\sinfluencedby(u,v)] \gets 1 - \prob[\overbar{\sinfluencedby(u,v)}]$ \Comment{Finalize the probability of $\sinfluencedby(u,v)$}
        \State $\prob[\overbar{\qinfluencedby(u,v)}] \gets \prob[\overbar{\sinfluencedby(u,v)}]$ \Comment{Probability of the complement of $\qinfluencedby(u,v)$}
        \If {$\level{u} <$ MAX\_LEVEL}
        \ForEach {$(v, w) \in E$}
        \If {$\state{w} \neq$ \textbf{processed}}
        \If {$\state{w} = $ \textbf{unprocessed} \textbf{and} $w \not \in S_i$}
        \State Enqueue$(Q, w)$
        \State $\state{w} \gets$ \textbf{in-process}
        \State $\level{w} \gets \level{v} + 1$
        \EndIf
        \State $\prob[\overbar{\sinfluencedby(u,w)}] \gets \prob[\overbar{\sinfluencedby(u,w)}] \cdot (1-\edgeprob(v,w) \cdot \prob[\sinfluencedby(u,v)]) $
        \Else
        \State $\prob[\overbar{\qinfluencedby(u,v)}] \gets \prob[\overbar{\qinfluencedby(u,v)}] \cdot (1-\edgeprob(v,w) \cdot \prob[\sinfluencedby(u,v)])$
        \EndIf
        \EndFor
        \EndIf
        \State $\state{w} \gets$ \textbf{processed}
        \EndWhile
        \ForEach {$v \in V$}
        \State $\prob[\qinfluencedby(u,v)] \gets 1 - \prob[\overbar{\qinfluencedby(u,v)}]$ \Comment{Finalize the probability of $\qinfluencedby(u,v)$}
        \State $\prob[\qinfluencedby(S_{i-1}\cup \{u\},v)] \gets \prob[\qinfluencedby(S_{i-1},v)] + (1 - \prob[\qinfluencedby(S_{i-1},v)]) \cdot \prob[\qinfluencedby(u,v)]$ \Comment{Finalize the probability of $\qinfluencedby(u,v)$}
        \EndFor
        \State \Return $\prob[\qinfluencedby(S_{i-1}\cup \{u\},v)]$ for each $v \in V$
    \end{algorithmic}
\end{algorithm}

\section{Experiments}\label{sec-experiments}
In this section we evaluate performance and efficiency of our algorithms and some previous algorithms on several real-world data-sets. We also present our experiments for selecting an appropriate value for $T$ in AAPC.

\subsection{Experiment Settings}
\textbf{Datasets.} We evaluate our implementation on three real-world datasets. Two of them are collaboration networks from e-print arXiv \url{https://arxiv.org}, High Energy Physics-Phenomenology section and Astro Physics~\cite{snapnets} which are denoted HEP and AstroP respectively. Third one is Epinions~\cite{snapnets} which is a who-trust-whom online social network of a general consumer review site \url{http://epinions.com}.Some basic
statistics of these networks are shown in Table~\ref{tab_datasets_info}.

\begin{table}[H] 
	\centering
	\caption{Statistics of the three real-world networks.}
	\label{tab_datasets_info}
	\begin{tabular}{|c|c|c|c|}
		\hline
		Name & $n$ & $m$ &   Type\\
		\hline
		High Energy Physics (HEP) &  12008 &  118521 & undirected \\ \hline
		Astro Physics (AstroP) & 18772  &  198110 &  undirected\\ \hline
		Epinions & 75879  & 508837  &  directed \\ \hline
	\end{tabular}  
\end{table}

\textbf{Algorithms.}
We compare our two proposed algorithms AAPC and EAAPC with five algorithms that are listed in the following. All of these algorithms are implemented in C++ and run on Intel Core i7-7700K @ 4.2GHz with 32 GB of RAM.

\begin{itemize}
	\item AAPC, Our proposed algorithm that computes the activation probabilities in $T$ steps. As the values of $p$ are small, we consider $T=4$.
	
	\item EAPC, Our proposed efficient algorithm.
	
	\item PMIA, the maximum influence path heuristic proposed in \cite{chen2010scalableMIA}.
	\item TowHops, the hop-based algorithm of \cite{tang2018efficient} that computes activation probability of each vertex through two hops from it,s in-neighbors.
	
	\item Greedy algorithm proposed in \cite{chen2009efficient} with lazy-forward optimization of \cite{leskovec2007cost}. For each seed set we run the simulations 5000 times. 
	
	\item IMM is a rivers influence sampling method from \cite{tang2015influence} with approximation guarantee.
	\item  DSSA is a one of the fast algorithms with approximation guarantee that is proposed in \cite{nguyen2016stop}.
\end{itemize}

\textbf{Influence propagation model.}
Since our algorithms are proposed on the IC model, we consider the propagation probability of all the edges to the exact value of $p$. As we explained in section~\ref{sec-proposed} it is clear that our algorithms have better performance on small values of $p$. So in our experiments we consider value of $p$ equal to 0.001, 0.005 and 0.01.

\subsection{Accuracy Discussion}\label{sec-exper-accuray}
\begin{figure}[h]
	\centering
	\includegraphics[scale=0.3]{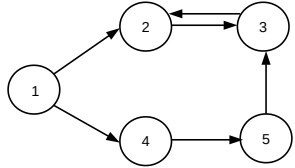}
	\caption{An example to find an appropriate value for $T$ in AAPC}\label{fig-apc-t}
\end{figure}

In section~\ref{sec-proposed} we mentioned that it is just enough to calculate the probabilities of the vertices until an appropriate time $t=T$. To have a good selection of $T$, we run the AAPC on graph of Figure~\ref{fig-apc-t} and report the activation probability of each vertex in Table~\ref{tab_apc_compare}.
Furthermore as we explained before, when there is any cycle in the graph, computation error of AAPC is less than previous works. So we also report the activation probabilities computed by Steady~State algorithm in the Table~\ref{tab_apc_compare}. To have a complete comparison, the exact probability of each vertex is also reported in colomn~\emph{exact prob}.  

All the probabilities in Table~\ref{tab_apc_compare} are computed by considering vertex~1 to be an active vertex and probability propagation of all the edges to be equal to $p=0.1$. 

According to the results, the variated probability of vertices 2 and 3 from $T=4$ to $T=5$ is a factor of $10^{-5}$ and in $T=6$ it is a factor of $10^{-6}$. As the $T$ increases this variated probability will become very small and therefore insignificant to the computations. So we offer $T=6$ as an appropriate value. When there are cycles in the graph the more computations not only add the more errors but also take more running-time. 

In Figure~\ref{fig-apc-t}, existence of cycle $(2,3,2)$ causes computing wrong probabilities for vertices 2 and 3. If we compare the probabilities of these vertices in step $T=6$ of AAPC with Steady~State we can see that AAPC has less computation errors. Indeed, AAPC is more accurate in activation probability computations. 

\begin{table}
	\caption{Comparison of activation probability computed by AAPC and Steady State on the graph of figure~\ref{fig-apc-t} to find an appropriate value for $T$} \label{tab_apc_compare}	
	\begin{center}
			\scriptsize
			\begin{tabular}{|c|c|c|c|c|c|c|c|c|c|}
				\hline
				& \multicolumn{7}{c|}{AAPC prob} & &\\
				\hline
				vertex & $T=0$ & $T=1$& $T=2$& $T=3$& $T=4$ & $T=5$ & $T=6$ & SteadyState prob & exact prop\\
				\hline
				1 & 1 & 1 & 1 & 1 & 1 & 1 & 1 & 1 & 1 \\
				\hline
				2 & 0 & 0.1 & 0.1 & 0.10081 & 0.100890 & 0.100897 & 0.100898  &  0.100998 & 0.10009 \\
				\hline
				3 & 0 & 0 & 0.01 & 0.01098 & 0.011068 &  0.011077  & 0.011078 & 0.011090 & 0.01099  \\
				\hline
				4 & 0 & 0.1 & 0.1 & 0.1 & 0.1 &  0.1 & 0.1  & 0.1 & 0.1 \\
				\hline
				5 & 0 & 0 & 0.01 & 0.01 & 0.01& 0.01 & 0.01 & 0.01 & 0.01 \\
				\hline
			\end{tabular}				
	\end{center}	
\end{table}

\subsection{Running Times and Influence Spread Analysis}
\begin{figure}[h]
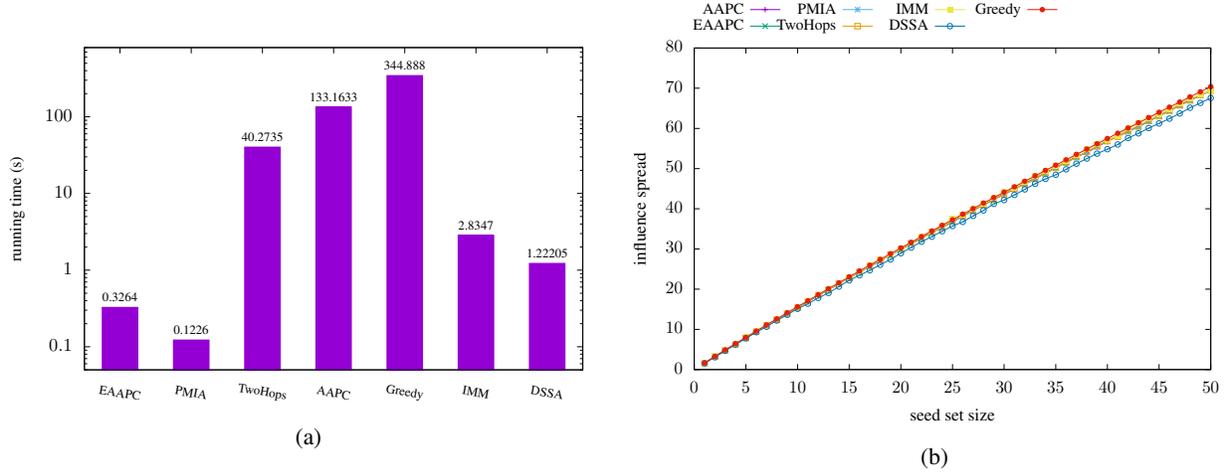

	\begin{subfigure}{.5\textwidth}\vspace{0.15cm}
			\resizebox{\textwidth}{!}{\input{charts/runtime_hep_001}}
		\caption{}\label{time-hep-0.001}
	\end{subfigure}
	\begin{subfigure}{.5\textwidth}
			\resizebox{\textwidth}{!}{\input{charts/hep_001}}	
		\caption{}\label{inf-hep-0.001}	
	\end{subfigure}
	\caption{Running times (a) and influence spreads (b) of algorithms on HEP under independent cascade model ($p=0.001$, $k=50$).}
	\label{chart_hep_0.001}
\end{figure}

Figures~\ref{chart_hep_0.001}, \ref{chart_hep_0.005} and \ref{chart_hep_0.01} show the running-time and influence spread of different algorithms on AstroP for $p=$ 0.001, 0.005 and 0.01, respectively. Figures~\ref{chart_astroph_0.001}--\ref{chart_epin_0.01} show the similar results on HEP and Epinions data-sets.

From running-times charts it can be seen that the AAPC is very faster than Greedy algorithm. At least it is 2.5 times faster than Greedy. Comparing running-time of Greedy for different values of $p$, shows that its running-time increases as the propagation probabilities of the edges increase while AAPC run independently.

In Figure~\ref{time-hep-0.001} both EAAPC and PMIA are very fast and their running-time is less than one seconds. But in Figures~\ref{time-hep-0.005} and \ref{time-hep-0.01} running-time of PMIA increases while EAAPC is still the fastest among all the algorithms.

Although in figures~\ref{inf-hep-0.001} and \ref{inf-hep-0.005} the influence spread of all the algorithms are close to each other, in Figure~\ref{inf-hep-0.01} for $p=0.01$ they are different. Greedy, IMM and DSSA have the most influence spread. After them AAPC and EAAPC have more influence spread than PMIA and TwoHops.
The reason is that AAPC and EAAPC consider more paths to calculate the influences in comparison to PMIA and TwoHops. Consequently, this lead to more accurate results.

The results of two other data-sets are similar to HEP and influence spread charts show that AAPC and EAAPC spread the influence as much as the other state-of-the-arts. They perform the same as Greedy, PMIA and DSSA which guarantee the approximation of the results. Furthermore AAPC has very less running-time than Greedy algorithms. The results show that EAAPC is the fastest among all the comparing algorithms and can be scaled on large-scale networks.

Another notable cases to explain is decreased influence spread of Greedy algorithm on Epinions in Figures~\ref{inf-epin-0.001}, \ref{inf-epin-0.005} and \ref{inf-epin-0.01}. The reason of this evidence is small number of influence simulations. Since Epinions is larger than the other two data-sets, it needs more simulations which make its running-time slower than it is. These figures also show that our EAAPC outperforms PMIA, because EAAPC considers most influential paths while PMIA just consider the most influential paths.

\begin{figure}[h]
	\begin{subfigure}{.5\textwidth}\vspace{0.15cm}
			\resizebox{\textwidth}{!}{\input{charts/runtime_hep_005}}
		\caption{}\label{time-hep-0.005}
	\end{subfigure}
	\begin{subfigure}{.5\textwidth}
			\resizebox{\textwidth}{!}{\input{charts/hep_005}}	
		\caption{}\label{inf-hep-0.005}	
	\end{subfigure}
	\caption{Running times (a) and influence spreads (b) of algorithms on HEP under independent cascade model ($p=0.005$, $k=50$).}
	\label{chart_hep_0.005}
\end{figure}

\begin{figure}[h]
	\begin{subfigure}{.5\textwidth}\vspace{0.15cm}
		\resizebox{\textwidth}{!}{\input{charts/runtime_hep_01}}		
		\caption{}\label{time-hep-0.01}
	\end{subfigure}
	\begin{subfigure}{.5\textwidth}		
		\resizebox{\textwidth}{!}{\input{charts/hep_01}}		
		\caption{}\label{inf-hep-0.01}	
	\end{subfigure}
	\caption{Running times (a) and influence spreads (b) of algorithms on HEP under independent cascade model ($p=0.01$, $k=50$).}
	\label{chart_hep_0.01}
\end{figure}

\begin{figure}[ht]
	\begin{subfigure}{.5\textwidth}\vspace{0.25cm}
		\resizebox{\textwidth}{!}{\input{charts/runtime_astroph_001}}
		\caption{}\label{time-astroph-0.001}
	\end{subfigure}
	\begin{subfigure}{.5\textwidth}
		\resizebox{\textwidth}{!}{\input{charts/astroph_001}}	
		\caption{}\label{inf-astroph-0.001}	
	\end{subfigure}
	\caption{Running times (a) and influence spreads (b) of algorithms on AstroP under independent cascade model ($p=0.001$, $k=50$).}
	\label{chart_astroph_0.001}
\end{figure}

\begin{figure}[ht]
	\begin{subfigure}{.5\textwidth}\vspace{0.15cm}		
		\resizebox{\textwidth}{!}{\input{charts/runtime_astroph_005}}		
		\caption{}\label{time-astroph-0.005}
	\end{subfigure}
	\begin{subfigure}{.5\textwidth}		
		\resizebox{\textwidth}{!}{\input{charts/astroph_005}}			
		\caption{}\label{inf-astroph-0.005}	
	\end{subfigure}
	\caption{Running times (a) and influence spreads (b) of algorithms on AstroP under independent cascade model ($p=0.005$, $k=50$).}
	\label{chart_astroph_0.005}
\end{figure}

\begin{figure}[ht]
	\begin{subfigure}{.5\textwidth}\vspace{0.15cm}		
		\resizebox{\textwidth}{!}{\input{charts/runtime_astroph_01}}		
		\caption{}\label{time-astroph-0.01}
	\end{subfigure}
	\begin{subfigure}{.5\textwidth}		
		\resizebox{\textwidth}{!}{\input{charts/astroph_01}}			
		\caption{}\label{inf-astroph-0.01}	
	\end{subfigure}
	\caption{Running times (a) and influence spreads (b) of algorithms on AstroP under independent cascade model ($p=0.01$, $k=50$).}
	\label{chart_astroph_0.01}
\end{figure}

\begin{figure}[ht]
	\begin{subfigure}{.5\textwidth}\vspace{0.15cm}
			\resizebox{\textwidth}{!}{\input{charts/runtime_epin_001}}
		\caption{}\label{time-epin-0.001}
	\end{subfigure}
	\begin{subfigure}{.5\textwidth}
			\resizebox{\textwidth}{!}{\input{charts/epin_001}}	
		\caption{}\label{inf-epin-0.001}	
	\end{subfigure}
	\caption{Running times (a) and influence spreads (b) of algorithms on Epinions under independent cascade model ($p=0.001$, $k=50$).}
	\label{chart_epin_0.001}
\end{figure}

\begin{figure}[ht]
	\begin{subfigure}{.5\textwidth}\vspace{0.15cm}
			\resizebox{\textwidth}{!}{\input{charts/runtime_epin_005}}
		\caption{}\label{time-epin-0.005}
	\end{subfigure}
	\begin{subfigure}{.5\textwidth}
			\resizebox{\textwidth}{!}{\input{charts/epin_005}}	
		\caption{}\label{inf-epin-0.005}	
	\end{subfigure}
		\caption{Running times (a) and influence spreads (b) of algorithms on Epinions under independent cascade model ($p=0.005$, $k=50$).}
	\label{chart_epin_0.005}
\end{figure}

\begin{figure}[ht]
	\begin{subfigure}{.5\textwidth}\vspace{0.15cm}		
		\resizebox{\textwidth}{!}{\input{charts/runtime_epin_01}}		
		\caption{}\label{time-epin-0.01}
	\end{subfigure}
	\begin{subfigure}{.5\textwidth}		
		\resizebox{\textwidth}{!}{\input{charts/epin_01}}			
		\caption{}\label{inf-epin-0.01}	
	\end{subfigure}
	\caption{Running times (a) and influence spreads (b) of algorithms on Epinions under independent cascade model ($p=0.01$, $k=50$).}
	\label{chart_epin_0.01}
\end{figure}
\section{Conclusion}\label{sec-conclusion}

In this paper we propose two greedy heuristics for influence maximization problem. 
Our proposed AAPC algorithm approximates the influence by using a novel analytic activation probability computation method which computes the activation probabilities more accurate than previous works.
The main idea of this algorithm is filling the gap between approximated and exact value of activation probability computation.
Our efficient EAAPC algorithm computes the activation probabilities through a subset of most influential paths to increase the accuracy of computations.
We examined our proposed algorithms on some real-world data-sets. The experiments show that our algorithms spread the influence as good as algorithms which guarantee the approximation of the results. 
In addition, our EAAPC has much improvements on the running-time so it can be easily scaled on large networks.
For future work, we are looking for more accurate techniques to calculate the influence spread of the vertices in faster running-time.

\bibliographystyle{plain}
\bibliography{references}
\end{document}